\documentclass[aip,apl]{revtex4-1}

\usepackage[version=3]{mhchem} 
\usepackage[utf8]{inputenc}
\usepackage{setspace}
\usepackage{amsmath}
\usepackage{amssymb}
\usepackage{csquotes}
\usepackage{graphicx}
\usepackage{subfig}
\usepackage{minibox}
\usepackage{wrapfig}
\usepackage{array}
\usepackage{url}
\usepackage[T1]{fontenc}

\draft 

\begin{document}


\title{Thermal transport mapping in twisted double bilayer graphene}

\author{Roop Kumar Mech}
\affiliation {\small \textit IMCN, Université Catholique de Louvain (UCLouvain), 1348 Louvain-la-Neuve, Belgium}

\author{Alessandra Canetta}
\affiliation {\small \textit IMCN, Université Catholique de Louvain (UCLouvain), 1348 Louvain-la-Neuve, Belgium}

\author{Yubin Huang}
\affiliation {\small \textit IMCN, Université Catholique de Louvain (UCLouvain), 1348 Louvain-la-Neuve, Belgium}

\author{Sergio Gonzalez-Munoz}
\affiliation {\small \textit Physics Department, Lancaster University, Lancaster, UK}

\author{Khushboo Agarwal}
\affiliation {\small \textit Physics Department, Lancaster University, Lancaster, UK}

\author{Pauline de Crombrugghe}
\affiliation {\small \textit IMCN, Université Catholique de Louvain (UCLouvain), 1348 Louvain-la-Neuve, Belgium}

\author{Yuanzhuo Hong}
\affiliation {\small \textit Universit\'e Paris-Saclay, CNRS, Centre de Nanosciences et de Nanotechnologies (C2N), 91120 Palaiseau, France}

\author{Sambit Mohapatra }
\affiliation {\small \textit Universit\'e Paris-Saclay, CNRS, Centre de Nanosciences et de Nanotechnologies (C2N), 91120 Palaiseau, France}

\author{Kenji Watanabe}
\affiliation {\small \textit Research Center for Electronic and Optical Materials, National Institute for Materials Science, 1-1 Namiki, Tsukuba 305-0044, Japan}

\author{Takashi Taniguchi}
\affiliation {\small \textit Research Center for Materials Nanoarchitectonics, National Institute for Materials Science,  1-1 Namiki, Tsukuba 305-0044, Japan}

\author{Bernard Nysten}
\affiliation {\small \textit IMCN, Université Catholique de Louvain (UCLouvain), 1348 Louvain-la-Neuve, Belgium}

\author{Benoît Hackens}
\affiliation {\small \textit IMCN, Université Catholique de Louvain (UCLouvain), 1348 Louvain-la-Neuve, Belgium}

\author{Rebeca Ribeiro-Palau}
\affiliation {\small \textit Universit\'e Paris-Saclay, CNRS, Centre de Nanosciences et de Nanotechnologies (C2N), 91120 Palaiseau, France}

\author{Oleg Kolosov}
\email{o.kolosov@lancaster.ac.uk}
\affiliation {\small \textit Physics Department, Lancaster University, Lancaster, UK}

\author{Jean Spièce}
\email{jean.spiece@uclouvain.be}
\affiliation {\small \textit IMCN, Université Catholique de Louvain (UCLouvain), 1348 Louvain-la-Neuve, Belgium}

\author{Pascal Gehring}
\email{pascal.gehring@uclouvain.be}
\affiliation {\small \textit IMCN, Université Catholique de Louvain (UCLouvain), 1348 Louvain-la-Neuve, Belgium}

\date{\today}

\begin{abstract}
Two-dimensional (2D) materials have attracted significant interest due to their tunable physical properties when stacked into homo- and hetero-structures. Twisting adjacent layers introduces moiré patterns that strongly influence the material electronic and thermal behavior. In twisted graphene systems, the twist angle critically alters phonon transport, leading to reduced thermal conductivity compared to Bernal-stacked configurations. However, experimental investigations into thermal transport in twisted structures remain limited. Here, we study the local thermal properties of twisted double bilayer graphene (TDBG) using Scanning Thermal Microscopy (SThM). We find an increase in thermal resistance of $0.3 \pm 0.1 \times 10^6$~K\,W$^{-1}$ compared to untwisted bilayers, attributed to changes in both intrinsic thermal conductivity and the tip–sample interface. These results, supported by analytical modeling, provide new insight into thermal transport mechanisms in twisted 2D systems and offer a pathway toward thermal engineering in twistronic devices.

\end{abstract}


\maketitle 

In recent years, two-dimensional (2D) materials have gained increasing attention. The atomic layers in these materials are held together by weak van der Waals forces, making them easy to cleave and stack into new heterostructures\cite{Geim13}. Stacking different 2D layers can result in artificial systems with novel physical properties\cite{Sutter21, Li17, Wu22}. Moreover, by twisting the layers with respect to each other, the periodicity of the resulting lattice changes, forming a so-called moir\'e pattern\cite{He21a}. This twist angle becomes a crucial parameter for tuning the band structure\cite{Ren20} and the lattice symmetry of the new material\cite{Carr17, Henn21}. At small twist angles\cite{Kere19}, the mismatch between layers leads to atomic reconstruction, favoring energetically stable stacking configurations. This results in discrete stacking domains and domain walls\cite{Yoo19}, which are predicted to significantly affect the local thermal properties of twisted 2D systems.

The thermal conductivity of graphene is known to depend on the twist angle between its layers. Bernal-stacked (AB) graphene exhibits the highest thermal conductivity. Introducing a twist reduces this property: Li \textit{et al.}~\cite{li2014} measured a 25\% decrease at $34^\circ$, while Han \textit{et al.}~\cite{han2021} observed a 15\% decrease at $2^\circ$, both using optothermal Raman spectroscopy. These reductions are attributed to changes in low-frequency phonon transport and are supported by molecular dynamics simulations. Other studies~\cite{Nie2019, Wang2017, li2018, krisna2025low} confirm that both in-plane and cross-plane thermal conductivities decrease as the twist angle increases.

Despite its importance for device performance and stability, thermal transport in twisted graphene structures remains poorly explored experimentally. A quantitative understanding of how twist affects heat dissipation in graphene could lead to angle-engineered thermal interfaces and optimized thermal management in twistronic devices. However, experimental confirmation of these predictions is still lacking.

Here, we investigate the local thermal properties of twisted double bilayer graphene (TDBG) using Scanning Thermal Microscopy (SThM). We observe an increase in the thermal resistance of TDBG by $0.3 \pm 0.1 \times 10^6$ KW${}^{-1}$ compared to untwisted double bilayer graphene. This increase is attributed to changes in both thermal conductivity and tip–sample thermal interface. Our findings are supported by analytical modeling.

\begin{figure}[t]
\centering \includegraphics[width=0.5\textwidth]{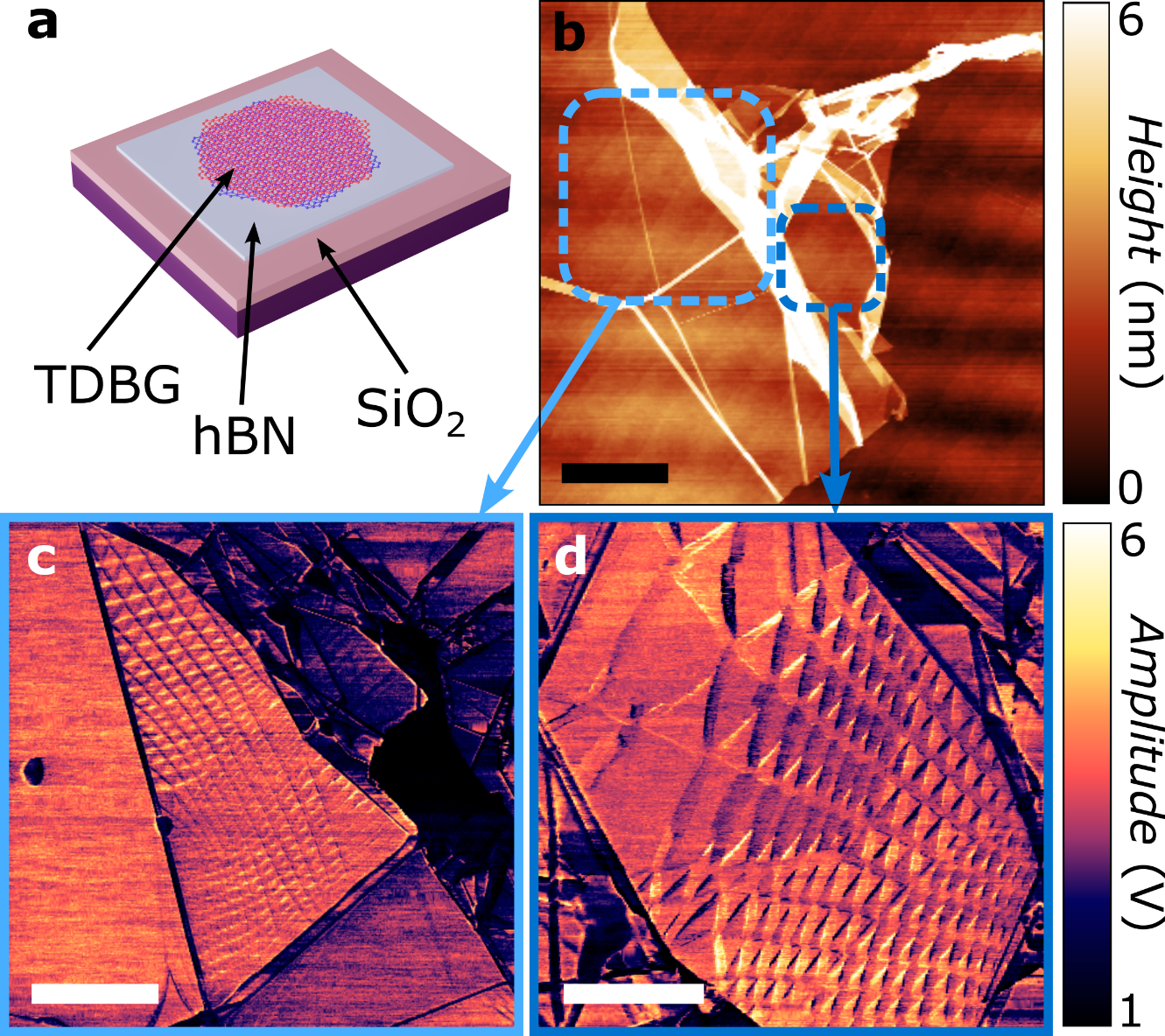}
\caption{(a) Scheme of the sample structure  consisting of a hBN flake acting as substrate on which two graphene bilayers where stacked at a small angle. The whole 2D stack is resting on a 300 nm SiO${}_2$ on Si wafer. (b) Topography map of the sample with the two regions showing large moiré patterns (highlighted by the dashed squares). Scale bar is 1$\mu$m. (c) PFM small scale image of the first region in (b). Scale bar is 250 nm (d) PFM image of the other moiré region. Scale is 100 nm.}
\label{fig1}
\end{figure}

Twisted double bilayer graphene consists of two AB-stacked bilayers placed on a 40 nm hexagonal boron nitride (hBN) substrate. A scheme of the full structure is shown in Figure~\ref{fig1}a. When stacked, a small twist angle $\theta$ ($<1^\circ$) is introduced between the bilayers, forming a moiré pattern of ABAB and ABCA domains, while suppressing the less stable ABBC regions\cite{Yoo19,Li21}. The TDBG was fabricated using a dry transfer method with a polycarbonate (PC) film (10\% concentration) supported by a polydimethylsiloxane bubble stamp. Two halves of a bilayer graphene, pre-cut using an atomic force microscope\cite{Li18}, were picked up sequentially with a relative twist angle of less than $1^\circ$. The final stack was transferred onto a hBN flake at room temperature without melting the PC\cite{Gade21}, minimizing contamination during the process.

The moiré wavelength $\lambda_m$ depends on the twist angle and is given by $\lambda_m = (a/2)/\sin(\theta/2)$, where $a$ is the graphene lattice constant\cite{Kim17}. Using piezoresponse force microscopy (PFM), we identified two regions with large moiré wavelengths (Figure~\ref{fig1}b). PFM is a useful tool to characterize twisted 2D systems due to their piezoelectric or flexoelectric responses\cite{McGil20}. In our PFM setup, an AC bias at the contact resonance frequency is applied between an electrically conductive AFM tip and the sample, generating a vertical electric field\cite{canetta2023quantifying,Hong21}. This field induces piezoelectric strain, which is detected through the AFM lever's torsion or deflection.

Figures~\ref{fig1}c and~\ref{fig1}d show PFM maps of the two regions. From these maps, we estimate effective twist angles between $0.07^\circ$ and $0.15^\circ$. Both regions display stretched triangular domains, which may indicate the presence of local strain or pinning at specific stacking sites\cite{Kere21}.

To investigate thermal transport in our 2D system, we use Scanning Thermal Microscopy, as described in previous studies\cite{evangeli2019,spiece2022}. SThM is a powerful technique that enables nanoscale mapping of heat dissipation in low-dimensional materials. Its operation is based on a sharp, microfabricated cantilever that functions both as a heater and a temperature sensor. The probe is raster-scanned across the sample surface using standard atomic force microscopy (AFM) feedback to control the tip–sample interaction.

Measurements were conducted under high vacuum conditions ($10^{-6}$ Torr) using a sharp probe (tip radius $\sim$30 nm), ensuring accurate thermal mapping without interference from air or water menisci\cite{evangeli2019,spiece2022}. We used commercially available doped silicon probes (Anasys Instruments, AN-300), which have a geometry similar to standard AFM probes, but with a cantilever composed of two highly doped, electrically conductive legs. These probes form part of a Wheatstone bridge circuit used to detect small changes in electrical resistance. A DC or AC voltage applied to the bridge heats the probe to a few tens of kelvin above room temperature. The output signal is amplified and calibrated to extract the probe temperature. Calibration details of both the thermal signal and probe resistance versus temperature can be found elsewhere\cite{spiece2018,spiece2019quantitative}.

\begin{figure}[t]
\centering \includegraphics[width=1\textwidth]{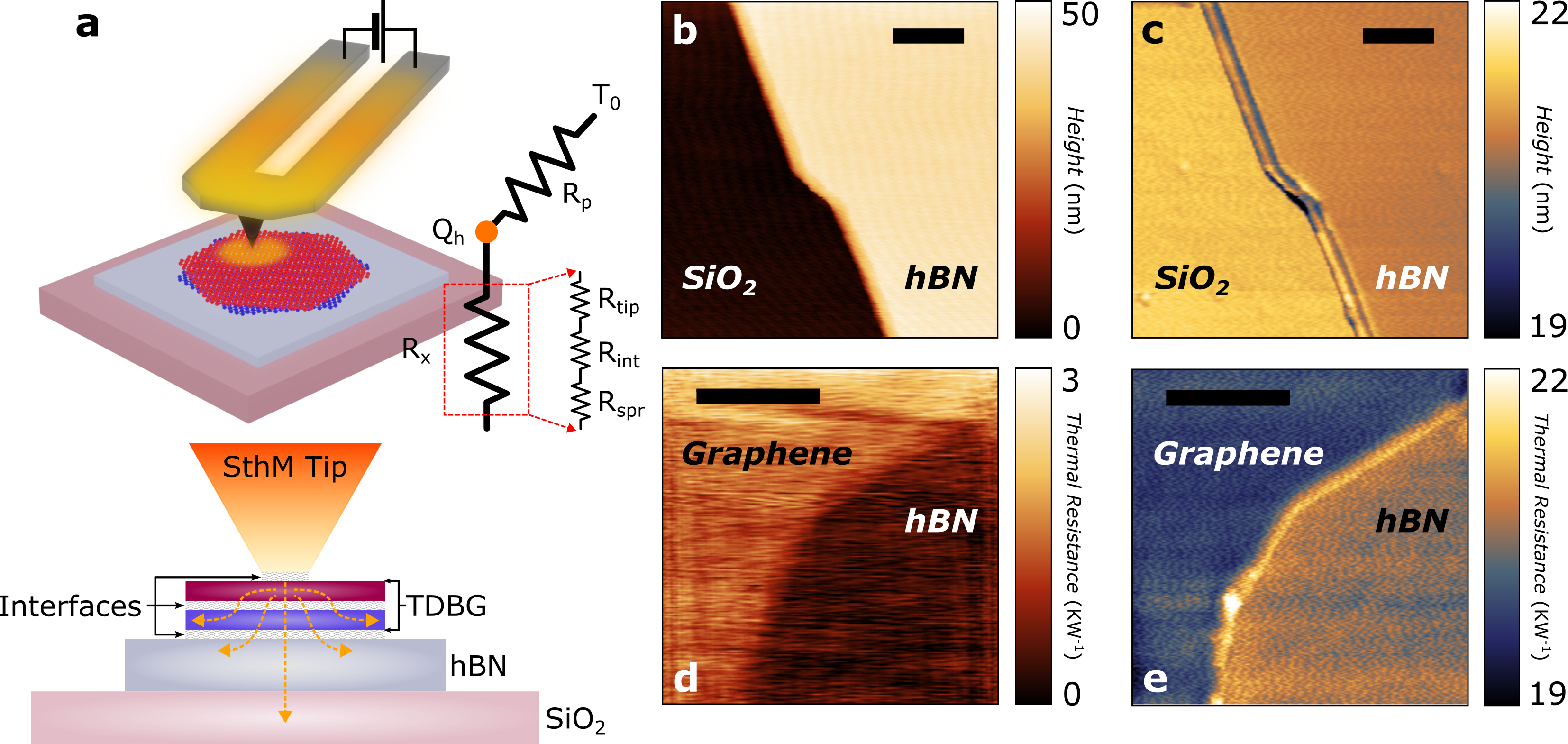}
\caption{(a) Schematic representation of the SThM probe and tip-sample system along with thermal resistances network (b,c) Topography and thermal resistance maps of the hBN flake on the silicon oxide substrate. (d,e) Topography and thermal resistance map of the step between hBN and 4 layers graphene. Lateral scale bars are 200 nm. Topography images and thermal resistance maps are in nm and KW${}^{-1}$, respectively.}
\label{fig2}
\end{figure}

Figure~\ref{fig2}a illustrates the working principle of SThM and its corresponding thermal resistance model. When the probe contacts a low-conductivity material, less heat flows into the sample, causing the probe temperature to rise. This temperature change alters the electrical resistance of the probe, which is detected via the bridge circuit. Under ambient conditions, the relation between the contact and non-contact voltages can be written as\cite{evangeli2019,spiece2022}:
\begin{equation}
    \frac{V_{nc}-V_c}{V_{nc}} = \frac{R_p}{R_p + R_X}
\end{equation}
where $V_{nc}$ and $V_c$ are the measured voltages with the probe out of contact and in contact with the sample, respectively. $R_p$ is the thermal resistance of the probe, and $R_X$ is the thermal resistance of the sample.

As shown later in Figure~\ref{fig2}a, the total measured thermal resistance $R_X$ consists of three components in series\cite{evangeli2019,spiece2022,huang2024violation}:
\begin{equation}
    R_X = R_{tip} + R_{int} + R_{spr} \label{eqRx}
\end{equation}
where $R_{tip}$ is the thermal resistance of the conical tip, $R_{int}$ is the tip–sample interface resistance, and $R_{spr}$ is the sample’s thermal spreading resistance. Proper interpretation of $R_X$ requires separating and analyzing each component.

Figures~\ref{fig2}b and \ref{fig2}c show topography and thermal resistance maps across a step between a silicon oxide (SiO$_x$) substrate and a 40 nm thick hBN flake. Due to hBN’s high thermal conductivity, its measured thermal resistance is lower than that of SiO$_x$. However, it is important to note that the apparent thermal resistance includes various contributions and artifacts, such as those from the tip–sample interface, surface properties, and topography\cite{gonzalez2023direct}.

Figures~\ref{fig2}d and \ref{fig2}e show a similar measurement across a boundary between hBN and a four-layer graphene region on hBN. When the probe transitions from hBN to graphene, we observe a clear drop in thermal resistance of $1 \pm 0.2 \times 10^6$~KW${}^{-1}$. While the difference in thermal conductivities—and therefore spreading resistances—between hBN and graphene may contribute to this contrast, it is not enough to fully account for the observed change. Since both materials are atomically flat and have high thermal conductivities, the main contribution must arise from variations in the tip–sample interface resistance.

The thermal interface resistance is given by $R_{int} = \frac{r_{int}^p}{\pi a^2}$ where $r_{int}^p$ is the thermal interface resistivity, and $a$ is the contact radius. We estimate $a$ by analyzing the variation in SThM and topography signals (see Supplementary Information for details), yielding a contact radius of $30.2 \pm 5$~nm. Using this and the observed thermal resistance drop, we calculate a change in interface resistivity of about $5 \pm 3$\%. This result confirms the high sensitivity of SThM to small variations in local heat transport. As we will show next, this sensitivity enables us to map thermal resistance differences between twisted graphene domains.

\begin{figure}[t]
\centering \includegraphics[width=0.75\textwidth]{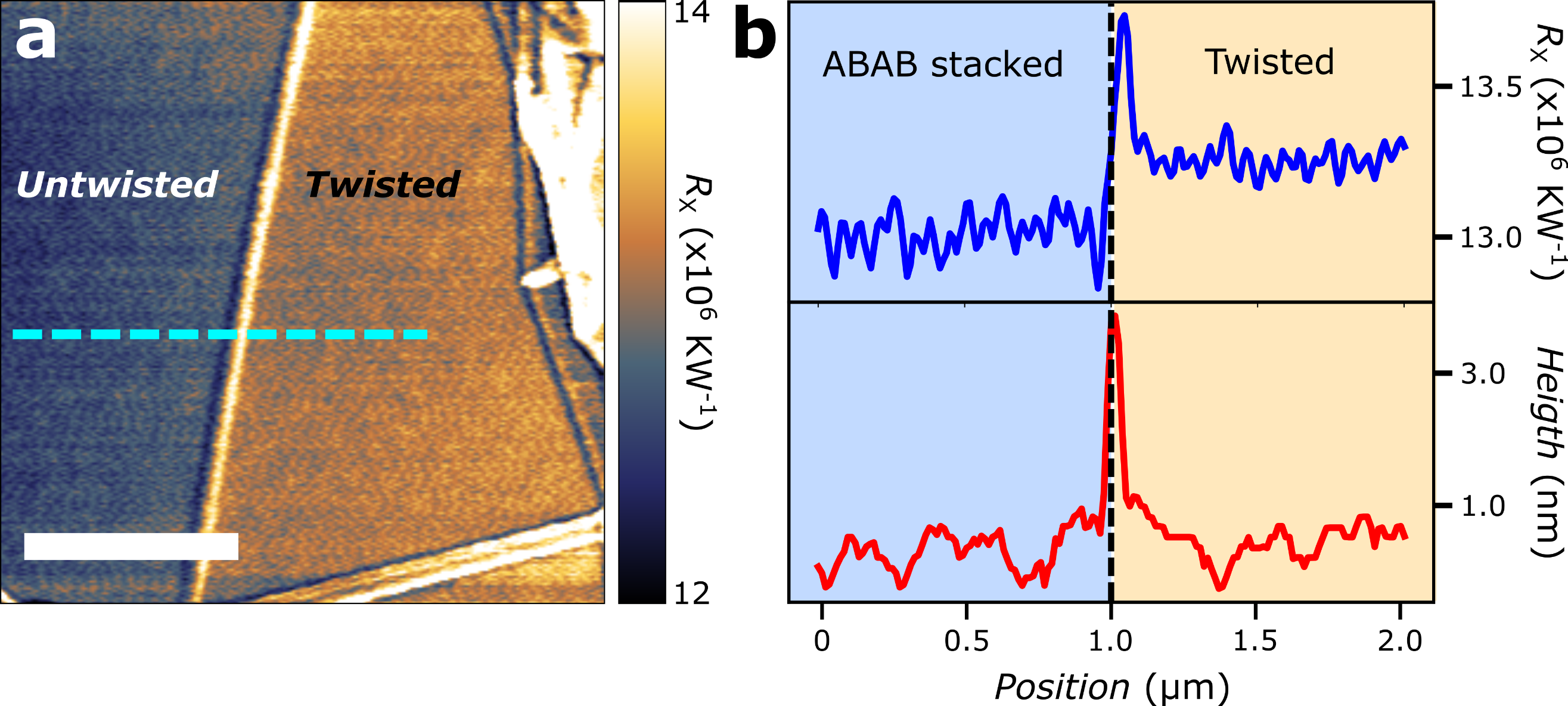} 
\caption{(a) Thermal resistance map showing higher resistance on the twisted region compared to the non-twisted one. Scale bar is 1 $\mu$m. (b) Thermal resistance and topography profiles taken across the transition between the two regions.} 
\label{fig3}
\end{figure}

Figure~\ref{fig3}a shows a Scanning Thermal Microscopy image of the first twisted graphene region. The scan was taken across the boundary between the twisted and non-twisted areas to allow direct comparison of the thermal signal. As seen in the image, the thermal resistance increases in the twisted area by $\Delta R_X = 0.3 \pm 0.1 \times 10^6$~KW${}^{-1}$. Topography and friction images reveal no visible differences between the two regions, confirming that the observed thermal contrast is not related to surface morphology or frictional properties.

A line profile extracted from the SThM image is shown in Figure~\ref{fig3}b, along with a simultaneously acquired topography profile. A clear step in thermal resistance is visible at the twisted–non-twisted boundary, while no notable change is observed in the topography, aside from a slight graphene ripple marking the interface. 

Next, we assess the relative contributions of the thermal interface resistance ($R_{int}$) and the thermal spreading resistance ($R_{spr}$) to the observed difference in measured resistance between twisted and non-twisted regions. Let $R_X^0$ and $R_X^t$ denote the measured thermal resistances in the non-twisted and twisted regions, respectively. From Eq.~\ref{eqRx}, and assuming the tip resistance $R_{tip}$ remains constant (as it is determined by the fixed probe geometry and apex size), the difference in thermal resistance is:
\begin{equation}\label{deltaRX}
    \Delta R_X = R_X^t - R_X^0 = R_{int}^t + R_{spr}^t - R_{int}^0 - R_{spr}^0
\end{equation}

The interface resistance ($R_{int}$) is often a dominant and uncertain factor in SThM measurements. It can be understood as the Kapitza resistance between the phonon systems of the probe and the sample. This resistance is influenced by material properties on both sides of the interface and any surface topography variations. In our case, the probe transitions from untwisted to twisted graphene. As shown in Figure~\ref{fig3}b, no significant topography change is detected at this boundary.

For graphene–SiO${}_2$ interfaces, reported thermal interface resistivity values range from $5.6 \times 10^{-9}$ to $7 \times 10^{-8}$~m${}^2$KW${}^{-1}$\cite{chen2009thermal,yasaei2017interfacial}. Based on our previously estimated contact radius, and assuming that the measured increase $\Delta R_X$ is solely due to changes in $R_{int}$, we estimate that the interface resistivity $r_{int}^{gr-SiO_2}$ increases by approximately 0.5–3\%, depending on the exact contact radius. 

Finally, we consider the contribution of the sample thermal spreading resistance ($R_{spr}$) to the overall measured variation. Given the high thermal conductivity of hBN, we assume it is well thermalized with the underlying silicon oxide substrate. Therefore, the relevant thermal spreading path is defined by the graphene–on–hBN stack, as illustrated in Figure~\ref{fig2}a. Analytical models for thermal spreading resistance in layered structures\cite{evangeli2019,spiece2022} can be applied here. These models include several key parameters (see Supplementary Information for full details): the contact radius $a$, the in-plane ($k_\parallel$) and cross-plane ($k_\perp$) thermal conductivities of graphene, the graphene–hBN interface resistivity ($r_{int}^{gr-hBN}$), and the thermal conductivity of hBN ($k_{hBN}$).

Next, we consider the in-plane and cross-plane thermal conductivities of the twisted graphene layers. Several studies\cite{wang2013high,zhang2017hexagonal,karak2021hexagonal} have shown that graphene partially recovers its exceptional thermal transport properties when supported by hBN. The thermal conductivity of suspended single-layer graphene has been measured\cite{balandin2008superior,pop2012thermal} to reach up to 3000~Wm${}^{-1}$K${}^{-1}$. When supported on high-quality boron nitride, reported values\cite{pop2012thermal} approach this suspended limit, typically ranging between 1000 and 2000~Wm${}^{-1}$K${}^{-1}$. For the cross-plane direction, bulk graphite exhibits\cite{pop2012thermal} a thermal conductivity of about 5~Wm${}^{-1}$K${}^{-1}$. 

Using these reported values as inputs, we find that to match the experimentally observed $\Delta R_X$ between the non-twisted and twisted regions, a tenfold reduction is required in either the in-plane or cross-plane thermal conductivity of the twisted graphene layer.

\begin{figure}[t]
\centering 
\includegraphics[width=0.5\textwidth]{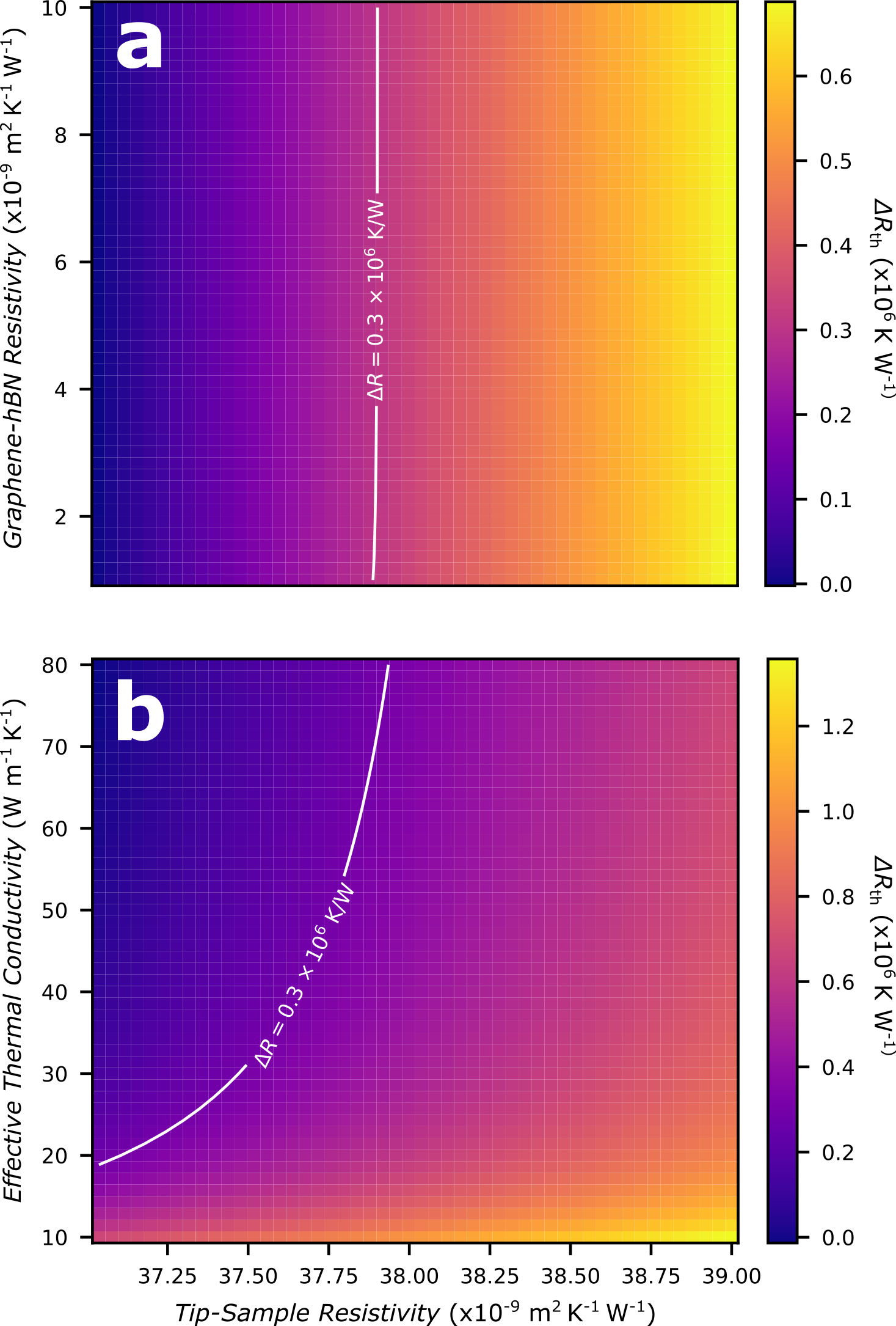}
\caption{Thermal resistance variation between twisted and non-twisted regions as a function of graphene-hBN resistivity and tip-sample resistivity (a) and effective thermal conductivity and tip-sample resistivity (b). On both panel, a white line is added to highlight the parameters' set matching the measured variation.} 
\label{fig4}
\end{figure}

To better understand the interplay between tip–sample interface resistance and thermal spreading resistance, we now analyze their relative contributions. In Figure~\ref{fig4}, we plot $\Delta R_X$ from Eq.~\ref{deltaRX} as a function of two sets of parameters: (a) tip–sample interface resistivity versus graphene–hBN interface resistivity, and (b) tip–sample resistivity versus effective thermal conductivity of the graphene stack. In both plots, we include a reference line corresponding to the experimentally measured $\Delta R_X$ of $0.3 \times 10^6$~KW${}^{-1}$.

From Figure~\ref{fig4}a, it is clear that variations in the graphene–hBN interface resistivity ($r_{int}^{gr-hBN}$) have a minor impact on the measured $\Delta R_X$. This is attributed to the high in-plane thermal conductivity of the TDBG, which effectively screens the contribution of the underlying interface. Conversely, Figure~\ref{fig4}b demonstrates that reducing the effective thermal conductivity of the graphene layer reproduces the observed $\Delta R_X$, similarly to increasing the tip–sample interface resistivity.

These results are consistent with prior reports on the effect of twisting on thermal transport in graphene systems, supporting the hypothesis that moiré engineering significantly alters phonon propagation in these materials.

In summary, we have investigated the local thermal transport properties of twisted double bilayer graphene supported on hexagonal boron nitride using scanning thermal microscopy. Our measurements reveal a measurable increase in thermal resistance in twisted regions compared to untwisted ones. Through analytical modeling, we attribute this change to a reduction in the effective thermal conductivity of the twisted stack, rather than interface or topographic effects. These results provide direct experimental evidence that twisting graphene layers—while widely used to tune electronic and optical properties—also significantly impacts heat dissipation. Our findings underscore the importance of considering twist angle as a design parameter for future two-dimensional devices, especially in applications where thermal management is critical.

\begin{acknowledgments}
We acknowledge financial support from the F.R.S.-FNRS of Belgium (FNRS-CQ-1.C044.21-SMARD, FNRS-CDR-J.0068.21-SMARD, FNRS-MIS-F.4523.22-TopoBrain, FNRS-CR-1.B.463.22-MouleFrits, FNRS-PDR-T.0029.22-Moir\'e), from the Federation Wallonie-Bruxelles through the ARC Grant No. 21/26-116 and from the EU (ERC-StG-10104144-MOUNTAIN). This project (40007563-CONNECT) has received funding from the FWO and F.R.S.-FNRS under the Excellence of Science (EOS) programme. This work was also partly supported by the FLAG-ERA Grant TATTOOS, through F.R.S.-FNRS PINT-MULTI Grant No. R 8010.19. We acknowledge the technical support from Bruker UK. The support of the European Union’s Horizon 2020 Research Project and Innovation Program–Graphene Flagship Core3 (No. 881603), EPSRC EP/V00767X/1 HiWiN project is fully appreciated. R.R.-P. acknowledges the Flag-Era JTC project TATTOOS (N$^{\circ}$ R.8010.19) and the Pathfinder project ``FLATS'' N$^{\circ}$ 101099139. Sample fabrication was done within the C2N micro nanotechnologies platforms and partly supported by the RENATECH network and the General Council of Essonne. K.W. and T.T. acknowledge support from the JSPS KAKENHI (Grant Numbers 21H05233 and 23H02052) , the CREST (JPMJCR24A5), JST and World Premier International Research Center Initiative (WPI), MEXT, Japan.
\end{acknowledgments}

\section*{DATA AVAILABILITY}
The data that support the findings of this study are available from the corresponding author upon reasonable request.

\bibliography{references}

\begin{thebibliography}{37}%
\makeatletter
\providecommand \@ifxundefined [1]{%
 \@ifx{#1\undefined}
}%
\providecommand \@ifnum [1]{%
 \ifnum #1\expandafter \@firstoftwo
 \else \expandafter \@secondoftwo
 \fi
}%
\providecommand \@ifx [1]{%
 \ifx #1\expandafter \@firstoftwo
 \else \expandafter \@secondoftwo
 \fi
}%
\providecommand \natexlab [1]{#1}%
\providecommand \enquote  [1]{``#1''}%
\providecommand \bibnamefont  [1]{#1}%
\providecommand \bibfnamefont [1]{#1}%
\providecommand \citenamefont [1]{#1}%
\providecommand \href@noop [0]{\@secondoftwo}%
\providecommand \href [0]{\begingroup \@sanitize@url \@href}%
\providecommand \@href[1]{\@@startlink{#1}\@@href}%
\providecommand \@@href[1]{\endgroup#1\@@endlink}%
\providecommand \@sanitize@url [0]{\catcode `\\12\catcode `\$12\catcode `\&12\catcode `\#12\catcode `\^12\catcode `\_12\catcode `\%12\relax}%
\providecommand \@@startlink[1]{}%
\providecommand \@@endlink[0]{}%
\providecommand \url  [0]{\begingroup\@sanitize@url \@url }%
\providecommand \@url [1]{\endgroup\@href {#1}{\urlprefix }}%
\providecommand \urlprefix  [0]{URL }%
\providecommand \Eprint [0]{\href }%
\providecommand \doibase [0]{http://dx.doi.org/}%
\providecommand \selectlanguage [0]{\@gobble}%
\providecommand \bibinfo  [0]{\@secondoftwo}%
\providecommand \bibfield  [0]{\@secondoftwo}%
\providecommand \translation [1]{[#1]}%
\providecommand \BibitemOpen [0]{}%
\providecommand \bibitemStop [0]{}%
\providecommand \bibitemNoStop [0]{.\EOS\space}%
\providecommand \EOS [0]{\spacefactor3000\relax}%
\providecommand \BibitemShut  [1]{\csname bibitem#1\endcsname}%
\let\auto@bib@innerbib\@empty
\bibitem [{\citenamefont {Geim}\ and\ \citenamefont {Grigorieva}(2013)}]{Geim13}%
  \BibitemOpen
  \bibfield  {author} {\bibinfo {author} {\bibfnamefont {A.~K.}\ \bibnamefont {Geim}}\ and\ \bibinfo {author} {\bibfnamefont {I.~V.}\ \bibnamefont {Grigorieva}},\ }\href@noop {} {\bibfield  {journal} {\bibinfo  {journal} {Nature}\ }\textbf {\bibinfo {volume} {499}},\ \bibinfo {pages} {419} (\bibinfo {year} {2013})}\BibitemShut {NoStop}%
\bibitem [{\citenamefont {Sutter}\ and\ \citenamefont {Sutter}(2021)}]{Sutter21}%
  \BibitemOpen
  \bibfield  {author} {\bibinfo {author} {\bibfnamefont {P.}~\bibnamefont {Sutter}}\ and\ \bibinfo {author} {\bibfnamefont {E.}~\bibnamefont {Sutter}},\ }\href@noop {} {\bibfield  {journal} {\bibinfo  {journal} {iScience}\ }\textbf {\bibinfo {volume} {24}},\ \bibinfo {pages} {103050} (\bibinfo {year} {2021})}\BibitemShut {NoStop}%
\bibitem [{\citenamefont {Li}, \citenamefont {Zhou},\ and\ \citenamefont {Zhang}(2017)}]{Li17}%
  \BibitemOpen
  \bibfield  {author} {\bibinfo {author} {\bibfnamefont {C.}~\bibnamefont {Li}}, \bibinfo {author} {\bibfnamefont {P.}~\bibnamefont {Zhou}}, \ and\ \bibinfo {author} {\bibfnamefont {D.~W.}\ \bibnamefont {Zhang}},\ }\href {\doibase 10.1088/1674-4926/38/3/031005} {\bibfield  {journal} {\bibinfo  {journal} {Journal of Semiconductors}\ }\textbf {\bibinfo {volume} {38}},\ \bibinfo {pages} {031005} (\bibinfo {year} {2017})}\BibitemShut {NoStop}%
\bibitem [{\citenamefont {Wu}\ \emph {et~al.}(2022)\citenamefont {Wu}, \citenamefont {Chen}, \citenamefont {Yang}, \citenamefont {Zhan}, \citenamefont {Ren},\ and\ \citenamefont {Li}}]{Wu22}%
  \BibitemOpen
  \bibfield  {author} {\bibinfo {author} {\bibfnamefont {X.}~\bibnamefont {Wu}}, \bibinfo {author} {\bibfnamefont {X.}~\bibnamefont {Chen}}, \bibinfo {author} {\bibfnamefont {R.}~\bibnamefont {Yang}}, \bibinfo {author} {\bibfnamefont {J.}~\bibnamefont {Zhan}}, \bibinfo {author} {\bibfnamefont {Y.}~\bibnamefont {Ren}}, \ and\ \bibinfo {author} {\bibfnamefont {K.}~\bibnamefont {Li}},\ }\href@noop {} {\bibfield  {journal} {\bibinfo  {journal} {Small}\ ,\ \bibinfo {pages} {2105877}} (\bibinfo {year} {2022})}\BibitemShut {NoStop}%
\bibitem [{\citenamefont {He}\ \emph {et~al.}(2021)\citenamefont {He}, \citenamefont {Zhou}, \citenamefont {Ye}, \citenamefont {Cho}, \citenamefont {Jeong}, \citenamefont {Meng},\ and\ \citenamefont {Wang}}]{He21a}%
  \BibitemOpen
  \bibfield  {author} {\bibinfo {author} {\bibfnamefont {F.}~\bibnamefont {He}}, \bibinfo {author} {\bibfnamefont {Y.}~\bibnamefont {Zhou}}, \bibinfo {author} {\bibfnamefont {Z.}~\bibnamefont {Ye}}, \bibinfo {author} {\bibfnamefont {S.-H.}\ \bibnamefont {Cho}}, \bibinfo {author} {\bibfnamefont {J.}~\bibnamefont {Jeong}}, \bibinfo {author} {\bibfnamefont {X.}~\bibnamefont {Meng}}, \ and\ \bibinfo {author} {\bibfnamefont {Y.}~\bibnamefont {Wang}},\ }\href {\doibase 10.1021/acsnano.0c10435} {\bibfield  {journal} {\bibinfo  {journal} {ACS Nano}\ }\textbf {\bibinfo {volume} {15}} (\bibinfo {year} {2021}),\ 10.1021/acsnano.0c10435}\BibitemShut {NoStop}%
\bibitem [{\citenamefont {Ren}\ \emph {et~al.}(2020)\citenamefont {Ren}, \citenamefont {Zhang}, \citenamefont {Liu},\ and\ \citenamefont {He}}]{Ren20}%
  \BibitemOpen
  \bibfield  {author} {\bibinfo {author} {\bibfnamefont {Y.-N.}\ \bibnamefont {Ren}}, \bibinfo {author} {\bibfnamefont {Y.}~\bibnamefont {Zhang}}, \bibinfo {author} {\bibfnamefont {Y.-W.}\ \bibnamefont {Liu}}, \ and\ \bibinfo {author} {\bibfnamefont {L.}~\bibnamefont {He}},\ }\href {\doibase 10.1088/1674-1056/abbbe2} {\bibfield  {journal} {\bibinfo  {journal} {Chinese Physics B}\ }\textbf {\bibinfo {volume} {29}},\ \bibinfo {pages} {117303} (\bibinfo {year} {2020})}\BibitemShut {NoStop}%
\bibitem [{\citenamefont {Carr}\ \emph {et~al.}(2017)\citenamefont {Carr}, \citenamefont {Massatt}, \citenamefont {Fang}, \citenamefont {Cazeaux}, \citenamefont {Luskin},\ and\ \citenamefont {Kaxiras}}]{Carr17}%
  \BibitemOpen
  \bibfield  {author} {\bibinfo {author} {\bibfnamefont {S.}~\bibnamefont {Carr}}, \bibinfo {author} {\bibfnamefont {D.}~\bibnamefont {Massatt}}, \bibinfo {author} {\bibfnamefont {S.}~\bibnamefont {Fang}}, \bibinfo {author} {\bibfnamefont {P.}~\bibnamefont {Cazeaux}}, \bibinfo {author} {\bibfnamefont {M.}~\bibnamefont {Luskin}}, \ and\ \bibinfo {author} {\bibfnamefont {E.}~\bibnamefont {Kaxiras}},\ }\href@noop {} {\bibfield  {journal} {\bibinfo  {journal} {PHYSICAL REVIEW B}\ }\textbf {\bibinfo {volume} {95}} (\bibinfo {year} {2017})}\BibitemShut {NoStop}%
\bibitem [{\citenamefont {Z.}\ and\ \citenamefont {S.}(2021)}]{Henn21}%
  \BibitemOpen
  \bibfield  {author} {\bibinfo {author} {\bibfnamefont {H.}~\bibnamefont {Z.}}\ and\ \bibinfo {author} {\bibfnamefont {K.}~\bibnamefont {S.}},\ }\href@noop {} {\bibfield  {journal} {\bibinfo  {journal} {Electron. Struct.}\ }\textbf {\bibinfo {volume} {3}} (\bibinfo {year} {2021})}\BibitemShut {NoStop}%
\bibitem [{\citenamefont {Kerelsky}\ \emph {et~al.}(2019)\citenamefont {Kerelsky}, \citenamefont {McGilly}, \citenamefont {Kennes}, \citenamefont {Xian}, \citenamefont {Yankowitz}, \citenamefont {Chen}, \citenamefont {Watanabe}, \citenamefont {Taniguchi}, \citenamefont {Hone}, \citenamefont {Dean} \emph {et~al.}}]{Kere19}%
  \BibitemOpen
  \bibfield  {author} {\bibinfo {author} {\bibfnamefont {A.}~\bibnamefont {Kerelsky}}, \bibinfo {author} {\bibfnamefont {L.~J.}\ \bibnamefont {McGilly}}, \bibinfo {author} {\bibfnamefont {D.~M.}\ \bibnamefont {Kennes}}, \bibinfo {author} {\bibfnamefont {L.}~\bibnamefont {Xian}}, \bibinfo {author} {\bibfnamefont {M.}~\bibnamefont {Yankowitz}}, \bibinfo {author} {\bibfnamefont {S.}~\bibnamefont {Chen}}, \bibinfo {author} {\bibfnamefont {K.}~\bibnamefont {Watanabe}}, \bibinfo {author} {\bibfnamefont {T.}~\bibnamefont {Taniguchi}}, \bibinfo {author} {\bibfnamefont {J.}~\bibnamefont {Hone}}, \bibinfo {author} {\bibfnamefont {C.}~\bibnamefont {Dean}},  \emph {et~al.},\ }\href@noop {} {\bibfield  {journal} {\bibinfo  {journal} {Nature}\ }\textbf {\bibinfo {volume} {572}},\ \bibinfo {pages} {95} (\bibinfo {year} {2019})}\BibitemShut {NoStop}%
\bibitem [{\citenamefont {Yoo}\ \emph {et~al.}(2019)\citenamefont {Yoo}, \citenamefont {Engelke}, \citenamefont {Carr},\ and\ \citenamefont {et~al.}}]{Yoo19}%
  \BibitemOpen
  \bibfield  {author} {\bibinfo {author} {\bibfnamefont {H.}~\bibnamefont {Yoo}}, \bibinfo {author} {\bibfnamefont {R.}~\bibnamefont {Engelke}}, \bibinfo {author} {\bibfnamefont {S.}~\bibnamefont {Carr}}, \ and\ \bibinfo {author} {\bibnamefont {et~al.}},\ }\href {\doibase https://doi.org/10.1038/s41563-019-0346-z} {\bibfield  {journal} {\bibinfo  {journal} {Nature Materials}\ }\textbf {\bibinfo {volume} {18}},\ \bibinfo {pages} {448} (\bibinfo {year} {2019})}\BibitemShut {NoStop}%
\bibitem [{\citenamefont {Li}\ \emph {et~al.}(2014)\citenamefont {Li}, \citenamefont {Ying}, \citenamefont {Chen}, \citenamefont {Nika}, \citenamefont {Cocemasov}, \citenamefont {Cai}, \citenamefont {Balandin},\ and\ \citenamefont {Chen}}]{li2014}%
  \BibitemOpen
  \bibfield  {author} {\bibinfo {author} {\bibfnamefont {H.}~\bibnamefont {Li}}, \bibinfo {author} {\bibfnamefont {H.}~\bibnamefont {Ying}}, \bibinfo {author} {\bibfnamefont {X.}~\bibnamefont {Chen}}, \bibinfo {author} {\bibfnamefont {D.~L.}\ \bibnamefont {Nika}}, \bibinfo {author} {\bibfnamefont {A.~I.}\ \bibnamefont {Cocemasov}}, \bibinfo {author} {\bibfnamefont {W.}~\bibnamefont {Cai}}, \bibinfo {author} {\bibfnamefont {A.~A.}\ \bibnamefont {Balandin}}, \ and\ \bibinfo {author} {\bibfnamefont {S.}~\bibnamefont {Chen}},\ }\href@noop {} {\bibfield  {journal} {\bibinfo  {journal} {Nanoscale}\ }\textbf {\bibinfo {volume} {6}},\ \bibinfo {pages} {13402} (\bibinfo {year} {2014})}\BibitemShut {NoStop}%
\bibitem [{\citenamefont {Han}\ \emph {et~al.}(2021)\citenamefont {Han}, \citenamefont {Nie}, \citenamefont {Gu}, \citenamefont {Liu}, \citenamefont {Chen}, \citenamefont {Ying}, \citenamefont {Wang}, \citenamefont {Cheng}, \citenamefont {Zhao},\ and\ \citenamefont {Chen}}]{han2021}%
  \BibitemOpen
  \bibfield  {author} {\bibinfo {author} {\bibfnamefont {S.}~\bibnamefont {Han}}, \bibinfo {author} {\bibfnamefont {X.}~\bibnamefont {Nie}}, \bibinfo {author} {\bibfnamefont {S.}~\bibnamefont {Gu}}, \bibinfo {author} {\bibfnamefont {W.}~\bibnamefont {Liu}}, \bibinfo {author} {\bibfnamefont {L.}~\bibnamefont {Chen}}, \bibinfo {author} {\bibfnamefont {H.}~\bibnamefont {Ying}}, \bibinfo {author} {\bibfnamefont {L.}~\bibnamefont {Wang}}, \bibinfo {author} {\bibfnamefont {Z.}~\bibnamefont {Cheng}}, \bibinfo {author} {\bibfnamefont {L.}~\bibnamefont {Zhao}}, \ and\ \bibinfo {author} {\bibfnamefont {S.}~\bibnamefont {Chen}},\ }\href@noop {} {\bibfield  {journal} {\bibinfo  {journal} {Applied Physics Letters}\ }\textbf {\bibinfo {volume} {118}},\ \bibinfo {pages} {193104} (\bibinfo {year} {2021})}\BibitemShut {NoStop}%
\bibitem [{\citenamefont {Nie}\ \emph {et~al.}(2019)\citenamefont {Nie}, \citenamefont {Zhao}, \citenamefont {Deng}, \citenamefont {Zhang},\ and\ \citenamefont {Du}}]{Nie2019}%
  \BibitemOpen
  \bibfield  {author} {\bibinfo {author} {\bibfnamefont {X.}~\bibnamefont {Nie}}, \bibinfo {author} {\bibfnamefont {L.}~\bibnamefont {Zhao}}, \bibinfo {author} {\bibfnamefont {S.}~\bibnamefont {Deng}}, \bibinfo {author} {\bibfnamefont {Y.}~\bibnamefont {Zhang}}, \ and\ \bibinfo {author} {\bibfnamefont {Z.}~\bibnamefont {Du}},\ }\href {\doibase https://doi.org/10.1016/j.ijheatmasstransfer.2019.03.130} {\bibfield  {journal} {\bibinfo  {journal} {International Journal of Heat and Mass Transfer}\ }\textbf {\bibinfo {volume} {137}},\ \bibinfo {pages} {161} (\bibinfo {year} {2019})}\BibitemShut {NoStop}%
\bibitem [{\citenamefont {Wang}, \citenamefont {Xie},\ and\ \citenamefont {Chen}(2017)}]{Wang2017}%
  \BibitemOpen
  \bibfield  {author} {\bibinfo {author} {\bibfnamefont {M.-H.}\ \bibnamefont {Wang}}, \bibinfo {author} {\bibfnamefont {Y.-E.}\ \bibnamefont {Xie}}, \ and\ \bibinfo {author} {\bibfnamefont {Y.-P.}\ \bibnamefont {Chen}},\ }\href {\doibase 10.1088/1674-1056/26/11/116503} {\bibfield  {journal} {\bibinfo  {journal} {Chinese Physics B}\ }\textbf {\bibinfo {volume} {26}},\ \bibinfo {pages} {116503} (\bibinfo {year} {2017})}\BibitemShut {NoStop}%
\bibitem [{\citenamefont {Li}\ \emph {et~al.}(2018{\natexlab{a}})\citenamefont {Li}, \citenamefont {Debnath}, \citenamefont {Tan}, \citenamefont {Su}, \citenamefont {Xu}, \citenamefont {Ge}, \citenamefont {Neupane},\ and\ \citenamefont {Lake}}]{li2018}%
  \BibitemOpen
  \bibfield  {author} {\bibinfo {author} {\bibfnamefont {C.}~\bibnamefont {Li}}, \bibinfo {author} {\bibfnamefont {B.}~\bibnamefont {Debnath}}, \bibinfo {author} {\bibfnamefont {X.}~\bibnamefont {Tan}}, \bibinfo {author} {\bibfnamefont {S.}~\bibnamefont {Su}}, \bibinfo {author} {\bibfnamefont {K.}~\bibnamefont {Xu}}, \bibinfo {author} {\bibfnamefont {S.}~\bibnamefont {Ge}}, \bibinfo {author} {\bibfnamefont {M.~R.}\ \bibnamefont {Neupane}}, \ and\ \bibinfo {author} {\bibfnamefont {R.~K.}\ \bibnamefont {Lake}},\ }\href@noop {} {\bibfield  {journal} {\bibinfo  {journal} {Carbon}\ }\textbf {\bibinfo {volume} {138}},\ \bibinfo {pages} {451} (\bibinfo {year} {2018}{\natexlab{a}})}\BibitemShut {NoStop}%
\bibitem [{\citenamefont {Krisna}, \citenamefont {Kawakami},\ and\ \citenamefont {Koshino}(2025)}]{krisna2025low}%
  \BibitemOpen
  \bibfield  {author} {\bibinfo {author} {\bibfnamefont {L.~P.}\ \bibnamefont {Krisna}}, \bibinfo {author} {\bibfnamefont {T.}~\bibnamefont {Kawakami}}, \ and\ \bibinfo {author} {\bibfnamefont {M.}~\bibnamefont {Koshino}},\ }\href@noop {} {\bibfield  {journal} {\bibinfo  {journal} {Journal of the Physical Society of Japan}\ }\textbf {\bibinfo {volume} {94}},\ \bibinfo {pages} {044602} (\bibinfo {year} {2025})}\BibitemShut {NoStop}%
\bibitem [{\citenamefont {Li}\ \emph {et~al.}(2021)\citenamefont {Li}, \citenamefont {Wang}, \citenamefont {Tang}, \citenamefont {Wang}, \citenamefont {Watanabe}, \citenamefont {Taniguchi}, \citenamefont {Gamelin}, \citenamefont {Cobden}, \citenamefont {Yankowitz}, \citenamefont {Xu},\ and\ \citenamefont {Li}}]{Li21}%
  \BibitemOpen
  \bibfield  {author} {\bibinfo {author} {\bibfnamefont {Y.}~\bibnamefont {Li}}, \bibinfo {author} {\bibfnamefont {X.}~\bibnamefont {Wang}}, \bibinfo {author} {\bibfnamefont {D.}~\bibnamefont {Tang}}, \bibinfo {author} {\bibfnamefont {X.}~\bibnamefont {Wang}}, \bibinfo {author} {\bibfnamefont {K.}~\bibnamefont {Watanabe}}, \bibinfo {author} {\bibfnamefont {T.}~\bibnamefont {Taniguchi}}, \bibinfo {author} {\bibfnamefont {D.~R.}\ \bibnamefont {Gamelin}}, \bibinfo {author} {\bibfnamefont {D.~H.}\ \bibnamefont {Cobden}}, \bibinfo {author} {\bibfnamefont {M.}~\bibnamefont {Yankowitz}}, \bibinfo {author} {\bibfnamefont {X.}~\bibnamefont {Xu}}, \ and\ \bibinfo {author} {\bibfnamefont {J.}~\bibnamefont {Li}},\ }\href {\doibase https://doi.org/10.1002/adma.202105879} {\bibfield  {journal} {\bibinfo  {journal} {Advanced Materials}\ }\textbf {\bibinfo {volume} {33}},\ \bibinfo {pages} {2105879} (\bibinfo {year} {2021})}\BibitemShut {NoStop}%
\bibitem [{\citenamefont {Li}\ \emph {et~al.}(2018{\natexlab{b}})\citenamefont {Li}, \citenamefont {Ying}, \citenamefont {Lyu}, \citenamefont {Deng}, \citenamefont {Wang}, \citenamefont {Taniguchi}, \citenamefont {Watanabe},\ and\ \citenamefont {Shi}}]{Li18}%
  \BibitemOpen
  \bibfield  {author} {\bibinfo {author} {\bibfnamefont {H.}~\bibnamefont {Li}}, \bibinfo {author} {\bibfnamefont {Z.}~\bibnamefont {Ying}}, \bibinfo {author} {\bibfnamefont {B.}~\bibnamefont {Lyu}}, \bibinfo {author} {\bibfnamefont {A.}~\bibnamefont {Deng}}, \bibinfo {author} {\bibfnamefont {L.}~\bibnamefont {Wang}}, \bibinfo {author} {\bibfnamefont {T.}~\bibnamefont {Taniguchi}}, \bibinfo {author} {\bibfnamefont {K.}~\bibnamefont {Watanabe}}, \ and\ \bibinfo {author} {\bibfnamefont {Z.}~\bibnamefont {Shi}},\ }\href {\doibase 10.1021/acs.nanolett.8b04166} {\bibfield  {journal} {\bibinfo  {journal} {Nano Letters}\ }\textbf {\bibinfo {volume} {18}} (\bibinfo {year} {2018}{\natexlab{b}}),\ 10.1021/acs.nanolett.8b04166}\BibitemShut {NoStop}%
\bibitem [{\citenamefont {Gadelha}\ \emph {et~al.}(2021)\citenamefont {Gadelha}, \citenamefont {Ohlberg}, \citenamefont {Santana}, \citenamefont {Silva}, \citenamefont {Lemos}, \citenamefont {Ornelas}, \citenamefont {Miranda}, \citenamefont {Nadas}, \citenamefont {Watanabe}, \citenamefont {Taniguchi}, \citenamefont {Rabelo}, \citenamefont {Venezuela}, \citenamefont {Medeiros-Ribeiro}, \citenamefont {Jorio}, \citenamefont {Cançado},\ and\ \citenamefont {Campos}}]{Gade21}%
  \BibitemOpen
  \bibfield  {author} {\bibinfo {author} {\bibfnamefont {A.}~\bibnamefont {Gadelha}}, \bibinfo {author} {\bibfnamefont {D.}~\bibnamefont {Ohlberg}}, \bibinfo {author} {\bibfnamefont {F.}~\bibnamefont {Santana}}, \bibinfo {author} {\bibfnamefont {E.}~\bibnamefont {Silva}}, \bibinfo {author} {\bibfnamefont {J.}~\bibnamefont {Lemos}}, \bibinfo {author} {\bibfnamefont {V.}~\bibnamefont {Ornelas}}, \bibinfo {author} {\bibfnamefont {D.}~\bibnamefont {Miranda}}, \bibinfo {author} {\bibfnamefont {R.}~\bibnamefont {Nadas}}, \bibinfo {author} {\bibfnamefont {K.}~\bibnamefont {Watanabe}}, \bibinfo {author} {\bibfnamefont {T.}~\bibnamefont {Taniguchi}}, \bibinfo {author} {\bibfnamefont {C.}~\bibnamefont {Rabelo}}, \bibinfo {author} {\bibfnamefont {P.}~\bibnamefont {Venezuela}}, \bibinfo {author} {\bibfnamefont {G.}~\bibnamefont {Medeiros-Ribeiro}}, \bibinfo {author} {\bibfnamefont {A.}~\bibnamefont {Jorio}}, \bibinfo {author} {\bibfnamefont {L.}~\bibnamefont {Cançado}}, \ and\ \bibinfo {author} {\bibfnamefont
  {L.}~\bibnamefont {Campos}},\ }\href {\doibase 10.1021/acsanm.0c03230} {\  (\bibinfo {year} {2021}),\ 10.1021/acsanm.0c03230}\BibitemShut {NoStop}%
\bibitem [{\citenamefont {Kim}\ \emph {et~al.}(2017)\citenamefont {Kim}, \citenamefont {DaSilva}, \citenamefont {Huang}, \citenamefont {Fallahazad}, \citenamefont {Larentis}, \citenamefont {Taniguchi}, \citenamefont {Watanabe}, \citenamefont {LeRoy}, \citenamefont {MacDonald},\ and\ \citenamefont {Tutuc}}]{Kim17}%
  \BibitemOpen
  \bibfield  {author} {\bibinfo {author} {\bibfnamefont {K.}~\bibnamefont {Kim}}, \bibinfo {author} {\bibfnamefont {A.}~\bibnamefont {DaSilva}}, \bibinfo {author} {\bibfnamefont {S.}~\bibnamefont {Huang}}, \bibinfo {author} {\bibfnamefont {B.}~\bibnamefont {Fallahazad}}, \bibinfo {author} {\bibfnamefont {S.}~\bibnamefont {Larentis}}, \bibinfo {author} {\bibfnamefont {T.}~\bibnamefont {Taniguchi}}, \bibinfo {author} {\bibfnamefont {K.}~\bibnamefont {Watanabe}}, \bibinfo {author} {\bibfnamefont {B.~J.}\ \bibnamefont {LeRoy}}, \bibinfo {author} {\bibfnamefont {A.~H.}\ \bibnamefont {MacDonald}}, \ and\ \bibinfo {author} {\bibfnamefont {E.}~\bibnamefont {Tutuc}},\ }\href {\doibase 10.1073/pnas.1620140114} {\bibfield  {journal} {\bibinfo  {journal} {Proceedings of the National Academy of Sciences}\ }\textbf {\bibinfo {volume} {114}},\ \bibinfo {pages} {3364} (\bibinfo {year} {2017})},\ \Eprint {http://arxiv.org/abs/https://www.pnas.org/doi/pdf/10.1073/pnas.1620140114}
  {https://www.pnas.org/doi/pdf/10.1073/pnas.1620140114} \BibitemShut {NoStop}%
\bibitem [{\citenamefont {McGilly}\ \emph {et~al.}(2020)\citenamefont {McGilly}, \citenamefont {Kerelsky}, \citenamefont {Finney}, \citenamefont {Shapovalov}, \citenamefont {Shih}, \citenamefont {Ghiotto}, \citenamefont {Zeng}, \citenamefont {Moore}, \citenamefont {Wu}, \citenamefont {Bai}, \citenamefont {Watanabe}, \citenamefont {Taniguchi}, \citenamefont {Stengel}, \citenamefont {Zhou}, \citenamefont {Hone}, \citenamefont {Zhu}, \citenamefont {Basov}, \citenamefont {Dean}, \citenamefont {Dreyer},\ and\ \citenamefont {Pasupathy}}]{McGil20}%
  \BibitemOpen
  \bibfield  {author} {\bibinfo {author} {\bibfnamefont {L.}~\bibnamefont {McGilly}}, \bibinfo {author} {\bibfnamefont {A.}~\bibnamefont {Kerelsky}}, \bibinfo {author} {\bibfnamefont {N.}~\bibnamefont {Finney}}, \bibinfo {author} {\bibfnamefont {K.}~\bibnamefont {Shapovalov}}, \bibinfo {author} {\bibfnamefont {E.-M.}\ \bibnamefont {Shih}}, \bibinfo {author} {\bibfnamefont {A.}~\bibnamefont {Ghiotto}}, \bibinfo {author} {\bibfnamefont {Y.}~\bibnamefont {Zeng}}, \bibinfo {author} {\bibfnamefont {S.}~\bibnamefont {Moore}}, \bibinfo {author} {\bibfnamefont {W.}~\bibnamefont {Wu}}, \bibinfo {author} {\bibfnamefont {Y.}~\bibnamefont {Bai}}, \bibinfo {author} {\bibfnamefont {K.}~\bibnamefont {Watanabe}}, \bibinfo {author} {\bibfnamefont {T.}~\bibnamefont {Taniguchi}}, \bibinfo {author} {\bibfnamefont {M.}~\bibnamefont {Stengel}}, \bibinfo {author} {\bibfnamefont {L.}~\bibnamefont {Zhou}}, \bibinfo {author} {\bibfnamefont {J.}~\bibnamefont {Hone}}, \bibinfo {author} {\bibfnamefont {X.}~\bibnamefont {Zhu}}, \bibinfo
  {author} {\bibfnamefont {D.}~\bibnamefont {Basov}}, \bibinfo {author} {\bibfnamefont {C.}~\bibnamefont {Dean}}, \bibinfo {author} {\bibfnamefont {C.}~\bibnamefont {Dreyer}}, \ and\ \bibinfo {author} {\bibfnamefont {A.}~\bibnamefont {Pasupathy}},\ }\href {\doibase 10.1038/s41565-020-0708-3} {\bibfield  {journal} {\bibinfo  {journal} {Nature Nanotechnology}\ }\textbf {\bibinfo {volume} {15}} (\bibinfo {year} {2020}),\ 10.1038/s41565-020-0708-3}\BibitemShut {NoStop}%
\bibitem [{\citenamefont {Canetta}\ \emph {et~al.}(2023)\citenamefont {Canetta}, \citenamefont {Gonzalez-Munoz}, \citenamefont {Nguyen}, \citenamefont {Agarwal}, \citenamefont {de~Picquendaele}, \citenamefont {Hong}, \citenamefont {Mohapatra}, \citenamefont {Watanabe}, \citenamefont {Taniguchi}, \citenamefont {Nysten} \emph {et~al.}}]{canetta2023quantifying}%
  \BibitemOpen
  \bibfield  {author} {\bibinfo {author} {\bibfnamefont {A.}~\bibnamefont {Canetta}}, \bibinfo {author} {\bibfnamefont {S.}~\bibnamefont {Gonzalez-Munoz}}, \bibinfo {author} {\bibfnamefont {V.-H.}\ \bibnamefont {Nguyen}}, \bibinfo {author} {\bibfnamefont {K.}~\bibnamefont {Agarwal}}, \bibinfo {author} {\bibfnamefont {P.~d.~C.}\ \bibnamefont {de~Picquendaele}}, \bibinfo {author} {\bibfnamefont {Y.}~\bibnamefont {Hong}}, \bibinfo {author} {\bibfnamefont {S.}~\bibnamefont {Mohapatra}}, \bibinfo {author} {\bibfnamefont {K.}~\bibnamefont {Watanabe}}, \bibinfo {author} {\bibfnamefont {T.}~\bibnamefont {Taniguchi}}, \bibinfo {author} {\bibfnamefont {B.}~\bibnamefont {Nysten}},  \emph {et~al.},\ }\href@noop {} {\bibfield  {journal} {\bibinfo  {journal} {Nanoscale}\ }\textbf {\bibinfo {volume} {15}},\ \bibinfo {pages} {8134} (\bibinfo {year} {2023})}\BibitemShut {NoStop}%
\bibitem [{\citenamefont {Hong}(2021)}]{Hong21}%
  \BibitemOpen
  \bibfield  {author} {\bibinfo {author} {\bibfnamefont {S.}~\bibnamefont {Hong}},\ }\href {\doibase 10.1063/5.0038744} {\bibfield  {journal} {\bibinfo  {journal} {Journal of Applied Physics}\ }\textbf {\bibinfo {volume} {129}},\ \bibinfo {pages} {051101} (\bibinfo {year} {2021})}\BibitemShut {NoStop}%
\bibitem [{\citenamefont {Kerelsky}\ \emph {et~al.}(2021)\citenamefont {Kerelsky}, \citenamefont {Rubio-Verdú}, \citenamefont {Xian}, \citenamefont {Kennes}, \citenamefont {Halbertal}, \citenamefont {Finney}, \citenamefont {Song}, \citenamefont {Turkel}, \citenamefont {Wang}, \citenamefont {Watanabe}, \citenamefont {Taniguchi}, \citenamefont {Hone}, \citenamefont {Dean}, \citenamefont {Basov}, \citenamefont {Rubio},\ and\ \citenamefont {Pasupathy}}]{Kere21}%
  \BibitemOpen
  \bibfield  {author} {\bibinfo {author} {\bibfnamefont {A.}~\bibnamefont {Kerelsky}}, \bibinfo {author} {\bibfnamefont {C.}~\bibnamefont {Rubio-Verdú}}, \bibinfo {author} {\bibfnamefont {L.}~\bibnamefont {Xian}}, \bibinfo {author} {\bibfnamefont {D.~M.}\ \bibnamefont {Kennes}}, \bibinfo {author} {\bibfnamefont {D.}~\bibnamefont {Halbertal}}, \bibinfo {author} {\bibfnamefont {N.}~\bibnamefont {Finney}}, \bibinfo {author} {\bibfnamefont {L.}~\bibnamefont {Song}}, \bibinfo {author} {\bibfnamefont {S.}~\bibnamefont {Turkel}}, \bibinfo {author} {\bibfnamefont {L.}~\bibnamefont {Wang}}, \bibinfo {author} {\bibfnamefont {K.}~\bibnamefont {Watanabe}}, \bibinfo {author} {\bibfnamefont {T.}~\bibnamefont {Taniguchi}}, \bibinfo {author} {\bibfnamefont {J.}~\bibnamefont {Hone}}, \bibinfo {author} {\bibfnamefont {C.}~\bibnamefont {Dean}}, \bibinfo {author} {\bibfnamefont {D.~N.}\ \bibnamefont {Basov}}, \bibinfo {author} {\bibfnamefont {A.}~\bibnamefont {Rubio}}, \ and\ \bibinfo {author} {\bibfnamefont {A.~N.}\
  \bibnamefont {Pasupathy}},\ }\href {\doibase 10.1073/pnas.2017366118} {\bibfield  {journal} {\bibinfo  {journal} {Proceedings of the National Academy of Sciences}\ }\textbf {\bibinfo {volume} {118}},\ \bibinfo {pages} {e2017366118} (\bibinfo {year} {2021})},\ \Eprint {http://arxiv.org/abs/https://www.pnas.org/doi/pdf/10.1073/pnas.2017366118} {https://www.pnas.org/doi/pdf/10.1073/pnas.2017366118} \BibitemShut {NoStop}%
\bibitem [{\citenamefont {Evangeli}\ \emph {et~al.}(2019)\citenamefont {Evangeli}, \citenamefont {Spiece}, \citenamefont {Sangtarash}, \citenamefont {Molina-Mendoza}, \citenamefont {Mucientes}, \citenamefont {Mueller}, \citenamefont {Lambert}, \citenamefont {Sadeghi},\ and\ \citenamefont {Kolosov}}]{evangeli2019}%
  \BibitemOpen
  \bibfield  {author} {\bibinfo {author} {\bibfnamefont {C.}~\bibnamefont {Evangeli}}, \bibinfo {author} {\bibfnamefont {J.}~\bibnamefont {Spiece}}, \bibinfo {author} {\bibfnamefont {S.}~\bibnamefont {Sangtarash}}, \bibinfo {author} {\bibfnamefont {A.~J.}\ \bibnamefont {Molina-Mendoza}}, \bibinfo {author} {\bibfnamefont {M.}~\bibnamefont {Mucientes}}, \bibinfo {author} {\bibfnamefont {T.}~\bibnamefont {Mueller}}, \bibinfo {author} {\bibfnamefont {C.}~\bibnamefont {Lambert}}, \bibinfo {author} {\bibfnamefont {H.}~\bibnamefont {Sadeghi}}, \ and\ \bibinfo {author} {\bibfnamefont {O.}~\bibnamefont {Kolosov}},\ }\href@noop {} {\bibfield  {journal} {\bibinfo  {journal} {Advanced Electronic Materials}\ }\textbf {\bibinfo {volume} {5}},\ \bibinfo {pages} {1900331} (\bibinfo {year} {2019})}\BibitemShut {NoStop}%
\bibitem [{\citenamefont {Spiece}\ \emph {et~al.}(2022)\citenamefont {Spiece}, \citenamefont {Sangtarash}, \citenamefont {Mucientes}, \citenamefont {Molina-Mendoza}, \citenamefont {Lulla}, \citenamefont {Mueller}, \citenamefont {Kolosov}, \citenamefont {Sadeghi},\ and\ \citenamefont {Evangeli}}]{spiece2022}%
  \BibitemOpen
  \bibfield  {author} {\bibinfo {author} {\bibfnamefont {J.}~\bibnamefont {Spiece}}, \bibinfo {author} {\bibfnamefont {S.}~\bibnamefont {Sangtarash}}, \bibinfo {author} {\bibfnamefont {M.}~\bibnamefont {Mucientes}}, \bibinfo {author} {\bibfnamefont {A.~J.}\ \bibnamefont {Molina-Mendoza}}, \bibinfo {author} {\bibfnamefont {K.}~\bibnamefont {Lulla}}, \bibinfo {author} {\bibfnamefont {T.}~\bibnamefont {Mueller}}, \bibinfo {author} {\bibfnamefont {O.}~\bibnamefont {Kolosov}}, \bibinfo {author} {\bibfnamefont {H.}~\bibnamefont {Sadeghi}}, \ and\ \bibinfo {author} {\bibfnamefont {C.}~\bibnamefont {Evangeli}},\ }\href@noop {} {\bibfield  {journal} {\bibinfo  {journal} {Nanoscale}\ }\textbf {\bibinfo {volume} {14}},\ \bibinfo {pages} {2593} (\bibinfo {year} {2022})}\BibitemShut {NoStop}%
\bibitem [{\citenamefont {Spiece}\ \emph {et~al.}(2018)\citenamefont {Spiece}, \citenamefont {Evangeli}, \citenamefont {Lulla}, \citenamefont {Robson}, \citenamefont {Robinson},\ and\ \citenamefont {Kolosov}}]{spiece2018}%
  \BibitemOpen
  \bibfield  {author} {\bibinfo {author} {\bibfnamefont {J.}~\bibnamefont {Spiece}}, \bibinfo {author} {\bibfnamefont {C.}~\bibnamefont {Evangeli}}, \bibinfo {author} {\bibfnamefont {K.}~\bibnamefont {Lulla}}, \bibinfo {author} {\bibfnamefont {A.}~\bibnamefont {Robson}}, \bibinfo {author} {\bibfnamefont {B.}~\bibnamefont {Robinson}}, \ and\ \bibinfo {author} {\bibfnamefont {O.}~\bibnamefont {Kolosov}},\ }\href@noop {} {\bibfield  {journal} {\bibinfo  {journal} {Journal of Applied Physics}\ }\textbf {\bibinfo {volume} {124}},\ \bibinfo {pages} {015101} (\bibinfo {year} {2018})}\BibitemShut {NoStop}%
\bibitem [{\citenamefont {Spi{\`e}ce}(2019)}]{spiece2019quantitative}%
  \BibitemOpen
  \bibfield  {author} {\bibinfo {author} {\bibfnamefont {J.}~\bibnamefont {Spi{\`e}ce}},\ }\href@noop {} {\emph {\bibinfo {title} {Quantitative mapping of nanothermal transport via Scanning Thermal Microscopy}}}\ (\bibinfo  {publisher} {Springer Nature},\ \bibinfo {year} {2019})\BibitemShut {NoStop}%
\bibitem [{\citenamefont {Huang}\ \emph {et~al.}(2024)\citenamefont {Huang}, \citenamefont {Spiece}, \citenamefont {Parker}, \citenamefont {Lee}, \citenamefont {Gogotsi},\ and\ \citenamefont {Gehring}}]{huang2024violation}%
  \BibitemOpen
  \bibfield  {author} {\bibinfo {author} {\bibfnamefont {Y.}~\bibnamefont {Huang}}, \bibinfo {author} {\bibfnamefont {J.}~\bibnamefont {Spiece}}, \bibinfo {author} {\bibfnamefont {T.}~\bibnamefont {Parker}}, \bibinfo {author} {\bibfnamefont {A.}~\bibnamefont {Lee}}, \bibinfo {author} {\bibfnamefont {Y.}~\bibnamefont {Gogotsi}}, \ and\ \bibinfo {author} {\bibfnamefont {P.}~\bibnamefont {Gehring}},\ }\href@noop {} {\bibfield  {journal} {\bibinfo  {journal} {ACS nano}\ }\textbf {\bibinfo {volume} {18}},\ \bibinfo {pages} {32491} (\bibinfo {year} {2024})}\BibitemShut {NoStop}%
\bibitem [{\citenamefont {Gonzalez-Munoz}\ \emph {et~al.}(2023)\citenamefont {Gonzalez-Munoz}, \citenamefont {Agarwal}, \citenamefont {Castanon}, \citenamefont {Kudrynskyi}, \citenamefont {Kovalyuk}, \citenamefont {Spi{\`e}ce}, \citenamefont {Kazakova}, \citenamefont {Patan{\`e}},\ and\ \citenamefont {Kolosov}}]{gonzalez2023direct}%
  \BibitemOpen
  \bibfield  {author} {\bibinfo {author} {\bibfnamefont {S.}~\bibnamefont {Gonzalez-Munoz}}, \bibinfo {author} {\bibfnamefont {K.}~\bibnamefont {Agarwal}}, \bibinfo {author} {\bibfnamefont {E.~G.}\ \bibnamefont {Castanon}}, \bibinfo {author} {\bibfnamefont {Z.~R.}\ \bibnamefont {Kudrynskyi}}, \bibinfo {author} {\bibfnamefont {Z.~D.}\ \bibnamefont {Kovalyuk}}, \bibinfo {author} {\bibfnamefont {J.}~\bibnamefont {Spi{\`e}ce}}, \bibinfo {author} {\bibfnamefont {O.}~\bibnamefont {Kazakova}}, \bibinfo {author} {\bibfnamefont {A.}~\bibnamefont {Patan{\`e}}}, \ and\ \bibinfo {author} {\bibfnamefont {O.~V.}\ \bibnamefont {Kolosov}},\ }\href@noop {} {\bibfield  {journal} {\bibinfo  {journal} {Advanced Materials Interfaces}\ }\textbf {\bibinfo {volume} {10}},\ \bibinfo {pages} {2300081} (\bibinfo {year} {2023})}\BibitemShut {NoStop}%
\bibitem [{\citenamefont {Chen}\ \emph {et~al.}(2009)\citenamefont {Chen}, \citenamefont {Jang}, \citenamefont {Bao}, \citenamefont {Lau},\ and\ \citenamefont {Dames}}]{chen2009thermal}%
  \BibitemOpen
  \bibfield  {author} {\bibinfo {author} {\bibfnamefont {Z.}~\bibnamefont {Chen}}, \bibinfo {author} {\bibfnamefont {W.}~\bibnamefont {Jang}}, \bibinfo {author} {\bibfnamefont {W.}~\bibnamefont {Bao}}, \bibinfo {author} {\bibfnamefont {C.}~\bibnamefont {Lau}}, \ and\ \bibinfo {author} {\bibfnamefont {C.}~\bibnamefont {Dames}},\ }\href@noop {} {\bibfield  {journal} {\bibinfo  {journal} {Applied Physics Letters}\ }\textbf {\bibinfo {volume} {95}} (\bibinfo {year} {2009})}\BibitemShut {NoStop}%
\bibitem [{\citenamefont {Yasaei}\ \emph {et~al.}(2017)\citenamefont {Yasaei}, \citenamefont {Foss}, \citenamefont {Karis}, \citenamefont {Behranginia}, \citenamefont {El-Ghandour}, \citenamefont {Fathizadeh}, \citenamefont {Olivares}, \citenamefont {Majee}, \citenamefont {Foster}, \citenamefont {Khalili-Araghi} \emph {et~al.}}]{yasaei2017interfacial}%
  \BibitemOpen
  \bibfield  {author} {\bibinfo {author} {\bibfnamefont {P.}~\bibnamefont {Yasaei}}, \bibinfo {author} {\bibfnamefont {C.~J.}\ \bibnamefont {Foss}}, \bibinfo {author} {\bibfnamefont {K.}~\bibnamefont {Karis}}, \bibinfo {author} {\bibfnamefont {A.}~\bibnamefont {Behranginia}}, \bibinfo {author} {\bibfnamefont {A.~I.}\ \bibnamefont {El-Ghandour}}, \bibinfo {author} {\bibfnamefont {A.}~\bibnamefont {Fathizadeh}}, \bibinfo {author} {\bibfnamefont {J.}~\bibnamefont {Olivares}}, \bibinfo {author} {\bibfnamefont {A.~K.}\ \bibnamefont {Majee}}, \bibinfo {author} {\bibfnamefont {C.~D.}\ \bibnamefont {Foster}}, \bibinfo {author} {\bibfnamefont {F.}~\bibnamefont {Khalili-Araghi}},  \emph {et~al.},\ }\href@noop {} {\bibfield  {journal} {\bibinfo  {journal} {Advanced Materials Interfaces}\ }\textbf {\bibinfo {volume} {4}},\ \bibinfo {pages} {1700334} (\bibinfo {year} {2017})}\BibitemShut {NoStop}%
\bibitem [{\citenamefont {Wang}, \citenamefont {Huang},\ and\ \citenamefont {Lu}(2013)}]{wang2013high}%
  \BibitemOpen
  \bibfield  {author} {\bibinfo {author} {\bibfnamefont {X.}~\bibnamefont {Wang}}, \bibinfo {author} {\bibfnamefont {T.}~\bibnamefont {Huang}}, \ and\ \bibinfo {author} {\bibfnamefont {S.}~\bibnamefont {Lu}},\ }\href@noop {} {\bibfield  {journal} {\bibinfo  {journal} {Applied Physics Express}\ }\textbf {\bibinfo {volume} {6}},\ \bibinfo {pages} {075202} (\bibinfo {year} {2013})}\BibitemShut {NoStop}%
\bibitem [{\citenamefont {Zhang}\ \emph {et~al.}(2017)\citenamefont {Zhang}, \citenamefont {Hu}, \citenamefont {Chen},\ and\ \citenamefont {Li}}]{zhang2017hexagonal}%
  \BibitemOpen
  \bibfield  {author} {\bibinfo {author} {\bibfnamefont {Z.}~\bibnamefont {Zhang}}, \bibinfo {author} {\bibfnamefont {S.}~\bibnamefont {Hu}}, \bibinfo {author} {\bibfnamefont {J.}~\bibnamefont {Chen}}, \ and\ \bibinfo {author} {\bibfnamefont {B.}~\bibnamefont {Li}},\ }\href@noop {} {\bibfield  {journal} {\bibinfo  {journal} {Nanotechnology}\ }\textbf {\bibinfo {volume} {28}},\ \bibinfo {pages} {225704} (\bibinfo {year} {2017})}\BibitemShut {NoStop}%
\bibitem [{\citenamefont {Karak}\ \emph {et~al.}(2021)\citenamefont {Karak}, \citenamefont {Paul}, \citenamefont {Negi}, \citenamefont {Poojitha}, \citenamefont {Srivastav}, \citenamefont {Das},\ and\ \citenamefont {Saha}}]{karak2021hexagonal}%
  \BibitemOpen
  \bibfield  {author} {\bibinfo {author} {\bibfnamefont {S.}~\bibnamefont {Karak}}, \bibinfo {author} {\bibfnamefont {S.}~\bibnamefont {Paul}}, \bibinfo {author} {\bibfnamefont {D.}~\bibnamefont {Negi}}, \bibinfo {author} {\bibfnamefont {B.}~\bibnamefont {Poojitha}}, \bibinfo {author} {\bibfnamefont {S.~K.}\ \bibnamefont {Srivastav}}, \bibinfo {author} {\bibfnamefont {A.}~\bibnamefont {Das}}, \ and\ \bibinfo {author} {\bibfnamefont {S.}~\bibnamefont {Saha}},\ }\href@noop {} {\bibfield  {journal} {\bibinfo  {journal} {ACS Applied Nano Materials}\ }\textbf {\bibinfo {volume} {4}},\ \bibinfo {pages} {1951} (\bibinfo {year} {2021})}\BibitemShut {NoStop}%
\bibitem [{\citenamefont {Balandin}\ \emph {et~al.}(2008)\citenamefont {Balandin}, \citenamefont {Ghosh}, \citenamefont {Bao}, \citenamefont {Calizo}, \citenamefont {Teweldebrhan}, \citenamefont {Miao},\ and\ \citenamefont {Lau}}]{balandin2008superior}%
  \BibitemOpen
  \bibfield  {author} {\bibinfo {author} {\bibfnamefont {A.~A.}\ \bibnamefont {Balandin}}, \bibinfo {author} {\bibfnamefont {S.}~\bibnamefont {Ghosh}}, \bibinfo {author} {\bibfnamefont {W.}~\bibnamefont {Bao}}, \bibinfo {author} {\bibfnamefont {I.}~\bibnamefont {Calizo}}, \bibinfo {author} {\bibfnamefont {D.}~\bibnamefont {Teweldebrhan}}, \bibinfo {author} {\bibfnamefont {F.}~\bibnamefont {Miao}}, \ and\ \bibinfo {author} {\bibfnamefont {C.~N.}\ \bibnamefont {Lau}},\ }\href@noop {} {\bibfield  {journal} {\bibinfo  {journal} {Nano letters}\ }\textbf {\bibinfo {volume} {8}},\ \bibinfo {pages} {902} (\bibinfo {year} {2008})}\BibitemShut {NoStop}%
\bibitem [{\citenamefont {Pop}, \citenamefont {Varshney},\ and\ \citenamefont {Roy}(2012)}]{pop2012thermal}%
  \BibitemOpen
  \bibfield  {author} {\bibinfo {author} {\bibfnamefont {E.}~\bibnamefont {Pop}}, \bibinfo {author} {\bibfnamefont {V.}~\bibnamefont {Varshney}}, \ and\ \bibinfo {author} {\bibfnamefont {A.~K.}\ \bibnamefont {Roy}},\ }\href@noop {} {\bibfield  {journal} {\bibinfo  {journal} {MRS bulletin}\ }\textbf {\bibinfo {volume} {37}},\ \bibinfo {pages} {1273} (\bibinfo {year} {2012})}\BibitemShut {NoStop}%
\end{thebibliography}%

\end{document}